# Quantum mechanics and discontinuous motion of particles

Rui Qi

Institute of Electronics, Chinese Academy of Sciences

17 Zhongguancun Rd., Beijing, China

E-mail: rg@mail.ie.ac.cn

We discuss a new realistic interpretation of quantum mechanics based on discontinuous motion of particles. The historical and logical basis of discontinuous motion of particles is given. It proves that if there exists only one kind of physical reality---particles, then the realistic motion of particles described by quantum mechanics should be discontinuous motion. We further denote that protective measurement may provide a direct method to confirm the existence of discontinuous motion of particles.

# Introduction

As we know, what classical mechanics describes is continuous motion of particles. Then a natural question appears when we turn to quantum mechanics, i.e. which motion of particles does quantum mechanics describe? This is not an easy question. In fact, people have been arguing with each other about its solution since the founding of quantum mechanics[1-5]. Present realistic interpretations of quantum mechanics all assume that the particles still undergo the continuous motion. But in order to hold this assumption, they must assume another kind of physical reality besides particles. Hidden variables theory[2] assumes the physical existence of $\psi$-wave field, which provides the quantum potential to create the quantum behaviors of the classical particles undergoing continuous motion. Many-worlds interpretation[3] also assumes the physical existence of $\psi$-wave field. In every world the particles undergo the continuous motion. Stochastic interpretation[4] assumes the existence of the background fluctuation field, which is required to result in the Brown motion of classical particles for accounting for their quantum displays.

Even though the present realistic interpretations of quantum mechanics are all dualistic, it doesn't preclude the existence of a monistic realistic interpretation of quantum mechanics. In a recent paper[6], a theory of discontinuous motion of particles is presented. It only assumes one kind of physical reality---particles, and concludes that the bizarre quantum behavior of particles only results from the motion of particles itself. Concretely speaking, it is shown that the wave function in quantum mechanics is the very mathematical complex describing the discontinuous motion of particles in continuous space-time, and the simplest nonrelativistic evolution equation of such motion is the same as Schroedinger equation. Considering the fact that space-time may be essentially discrete when combining quantum theory and general relativity, the author further analyzes the discontinuous motion of particles in discrete space-time, and show that its evolution may naturally result in the dynamical collapse process of the wave function. These analyses strongly imply what quantum theory describes may be discontinuous motion of particles.

In this paper, we will further give the historical and logical basis of discontinuous motion of particles. We demonstrate that if there exists only one kind of physical reality---particles, then the realistic motion of particles described by quantum mechanics should be discontinuous motion. It is also shown that protective measurement[7][8] just provides a direct method to confirm the existence of discontinuous motion of particles.

The plan of this paper is as follows: In Sect. 2 we first present a historical analysis of three important concepts related to the understanding of quantum mechanics. We denote that the combination of the concept of reality of motion, particle concept and discontinuity concept may naturally result in the existence of the realistic discontinuous motion of particles. In Sect. 3 we give a logical analysis of the discontinuous motion of particles. We re-analyze the famous double-slit experiment, and show that if there exists only one kind of physical reality---particles, then the realistic motion of particles described by quantum mechanics should be discontinuous motion. In Sect. 4 we discuss the possible experimental confirmation of the discontinuous motion of particles. We show that protective measurement may provide a direct method to confirm the existence of discontinuous motion of particles. Conclusions are given in Sect. 5.

## Three important concepts

From a historical point of view, three important concepts will help to form a monastic interpretation of quantum mechanics based only on the physical reality---particles.

The first essential concept is the ontology concept widely adopted in interpreting quantum mechanics. It aims at finding the last reality behind the mysterious wave function. This concept can be traced back to Einstein, Schrödinger, and the followers Bohm, Bell et al. Especially in recent years, the protective measurement presented by Aharonov et al[7][8] has not only consolidated this concept from the inside of quantum mechanics, but also provided the experimental method to confirm it. This advance has indicated that the reality underlying the wave function is more close to us than ever, and it may be the time to disclose the whole quantum mystery now.

The second is the particle concept hold by Born's widely-accepted probability interpretation and nearly most following interpretations of quantum mechanics. According to the particle concept, the reality described by the wave function in quantum mechanics is essentially a particle. It is in only one position in space at any instant, and possesses the qualities such as mass and charge etc. The most predominant character of this concept is that it not only can provide the particle picture in quantum mechanics, but also can be extended to describe the field picture in quantum field theory when considering special relativity. Besides, it also provides a precondition to further study the notorious collapse process of the wave function as one kind of objective process.

The third is the renunciation of classical continuous motion. It is widely demonstrated and accepted by physicists that the particle can not pass through only one slit in double-slit experiment, and sometimes they even say that the particle 'passes through both slits' or 'in two positions at the same time'. Although this kind of description is too vague to form a strict definition about the motion undergone by the particle, which may be extremely different from classical continuous motion, the renunciation of classical continuous motion undoubtedly sheds light on the road to interpret the wave function in terms of one kind of new motion of particles, and the vague description 'in two positions at the same time' indeed grasps something real and moves a peg along this road.

As we know, people have been studying continuous motion of particle since the founding of science, and taking it for granted that it is the only possible motion mode in Nature. Now, the combination of ontology concept, particle concept and the renunciation of continuous motion may naturally result in the existence of one kind of discontinuous motion of particles, which may further provide the basis of a monistic realistic interpretation of quantum mechanics.

## Another logical possibility

Present realistic interpretations of quantum mechanics all assume two kinds of reality---particles and field. Here we consider another logical possibility, namely there exists only one kind of reality---particles.

Even though people haven't known what does quantum mechanics describe yet, they indeed know what quantum mechanics does not describe. It is well known that, if there exist only particles, then quantum mechanics doesn't describe continuous motion of particles, or we can say, what quantum mechanics describes is not continuous motion of particles. Here as an example, let's have a look at the well-known double-slit experiment, and see why the motion described by quantum mechanics is not continuous motion of particles.

In the usual double-slit experiment, the single particle such as photon is emitted from the source one after the other, and then passes through the two slits to arrive at the screen. In this way, when a large number of particles reach the screen, they form the double-slit interference pattern. Now we will demonstrate that this experiment clearly reveals what quantum mechanics describes is not continuous motion of particles. If the motion of particle is continuous, then the particle can only pass through one of the two slits, and it is not influenced by the other slit in each experiment. Thus it is evident that the double-slit interference pattern will be the same as the direct mixture of two one-slit patterns, each of which is formed by opening each of the two slits, since the passing process of each particle in double-slit experiment is exactly the same as that in one of the two one-slit experiments. But quantum mechanics predicts that there exist obvious differences between the interference patterns of the above two situations, and all known experiments coincide with the prediction. Thus the motion of particle described by quantum mechanics can't be continuous, and the particle must pass through both slits in some unusual way during passing through the two slits.

Now there appears a simple but subtle question, i.e. if the motion of the particles described by quantum mechanics is not continuous, then which form is it? If there exist only particles, and the objective motion picture of the particles can't be essentially rejected, then the motion of particles should be discontinuous. This is an inevitable logical conclusion. As we think, this answer is more direct and natural, since classical mechanics describes continuous motion, then correspondingly quantum mechanics, which is essentially different from classical mechanics, will describe another different kind of motion, namely discontinuous motion. But the answer seems very bizarre, and we have never learned the discontinuous motion. In the following, let's be close to it and grasp it.

## A theory of discontinuous motion of particles

In a recent paper[6], a theory of discontinuous motion of particles is presented. It concludes that the bizarre quantum behavior of particles results only from the motion of particles itself. In this section, we will briefly introduce the theory of discontinuous motion of particles, which strongly imply what quantum theory describes is discontinuous motion of particles.

First, the strict definition of the discontinuous motion of particles is given using three presuppositions about the relation between physical motion and mathematical point set. They are the basic conceptions and correspondence rules needed before the discontinuous motion of particles in continuous space-time is discussed.

(1). Time and space in which the particle moves are both continuous.

(2). The moving particle is represented by one point in time and space.

(3). The discontinuous motion of particle is represented by the dense point set in time and space.

The first presupposition defines the continuity of space-time. The second one defines the existent form of particle in time and space. The last one defines the discontinuous motion of particle using the mathematical point set.

Then the mathematical description of the discontinuous motion of particle is given using the point set theory. It is shown that the proper description of the motion state of a particle undergoing the discontinuous motion, which forms a dense point set, is position measure density $r(x,t)$ and position measure flux density $j(x,t)$, which satisfy the measure conservation equation $\frac{\partial r(x,t)}{\partial t} + \frac{\partial j(x,t)}{\partial x} = 0$. The physical meaning of the position measure density $r(x,t)$ is that it represents the relative frequency of the particle appearing in the infinitesimal space interval $dx$ near the position $x$ during the infinitesimal interval $dt$ near the instant $t$. It can be measured by directly measuring the appearing probability of the particle in the above situation.

Furthermore, the descriptions of the motion of a single particle is naturally extended to the many particles situation. As to the motion state of N particles, the joint position measure density $r(x_1, x_2, ... x_N, t)$ can be defined according to point set theory, which represents the appearing probability of the situation, in which particle 1 is in position $x_1$, particle 2 is in position $x_2$, … and particle N is in position $x_N$. In a similar way, the joint position measure flux density $j(x_1, x_2, ... x_N, t)$ can be also defined through the joint measure conservation equation: $\frac{\partial r(x_1, x_2, ... x_N, t)}{\partial t} + \sum_{i=1}^{N} \frac{\partial j(x_1, x_2, ... x_N, t)}{\partial x_i} = 0$. It is stressed that the descriptions of the motion of many particles, namely the joint position measure density $r(x_1, x_2, ... x_N, t)$ and joint position measure flux density $j(x_1, x_2, ... x_N, t)$ are naturally defined in the 3N dimensional configure space, not in the real three-dimensional space.

After given the mathematical description of the discontinuous motion of particles, the possible evolution equations of the discontinuous motion of particles are further analyzed. First, the first motion principle of such motion is given. It asserts that during the free motion of particle, the momentum of the particle is invariant. It is denoted that, contrary to continuous motion, for the free particle with one constant momentum, its position will not be limited in the infinitesimal space interval $dx$, but spread throughout the whole space with the same position measure density.

Similar to the quantity position, the momentum (motion) state of a particle is further defined. It is described by the momentum measure density $f(p,t)$ and the momentum measure flux density $J(p,t)$. Their meanings are similar to those of position, and satisfy the similar measure conservation equation $\frac{\partial f(p,t)}{\partial t} + \frac{\partial J(p,t)}{\partial p} = 0$. Since the motion state of a particle is unique at any instant, there should exist a one-to-one relation between these two kinds of descriptions---position description $r(x,t)$, $j(x,t)$ and momentum description $f(p,t)$, $J(p,t)$, and this relation is irrelevant to the concrete motion state. The one-to-one relation is further obtained through some mathematical analyses on the essential symmetries involved in the discontinuous motion itself. The simplest form of the relation can be written as follows:

$$\psi(x,t) = \int_{-\infty}^{+\infty} j(p,t) e^{ipx - iEt} dp \quad \text{------ (1)}$$

This one-to-one relation holds true for any discontinuous motion state of particles.

On the basis of the one-to-one relation, the simplest nonrelativistic evolution law of the discontinuous motion of particles is worked out. The free evolution equation is:

$$i\hbar \frac{\partial \psi(x,t)}{\partial t} = -\frac{\hbar^2}{2m} \frac{\partial^2 \psi(x,t)}{\partial^2 x} \quad \text{------(2)}$$

The evolution equation including an outside potential is

$$i\hbar \frac{\partial \psi(x,t)}{\partial t} = -\frac{\hbar^2}{2m} \frac{\partial^2 \psi(x,t)}{\partial^2 x} + U(x,t)\psi(x,t) \quad \text{------(3)}$$

This indicates that the simplest nonrelativistic evolution equation of the discontinuous motion of particles is just Schrödinger equation in quantum mechanics. It is further denoted that there exists a one-to-one relation between $r(x,t)$, $j(x,t)$ and $\psi(x,t)$ when omitting the absolute phase, thus the state function $\psi(x,t)$ also provides a complete description of the discontinuous motion of particles.

## The meaning of the theory of discontinuous motion

The sameness between the simplest nonrelativistic evolution equation of the discontinuous motion of particles and the Schrödinger equation in quantum mechanics strongly suggests what quantum mechanics describes is discontinuous motion of particles. But before reaching the definite conclusion, we need to understand the meaning of the theory of discontinuous motion. This means we must talk about measurement.

One subtle problem is what happens during a measuring process. There exist only two possibilities: one is that the measuring process still satisfies the above evolution equation of discontinuous motion or Schrödinger equation, the linear superposition of the wave function can hold all through. This possibility corresponds to the present many worlds interpretation of quantum mechanics; the other is that the measuring process doesn't satisfy the above evolution equation of discontinuous motion or Schrödinger equation, the linear superposition of the wave function is destroyed due to some unknown causes. The resulting process is called the dynamical collapse of wave function, and the corresponding theory is generally named as revised quantum dynamics. Certainly, the above two possibilities can be tested in experiments, but unfortunately it is very difficult to distinguish them using present technology. In the following we will mainly give some theoretical considerations about them.

As to the first possibility, the discontinuous motion of particles provides the corresponding physical reality in continuous space-time for many worlds interpretation. The particle discontinuously moves throughout all the parallel worlds during very small time interval, or even infinitesimal time interval, and this objectively and clearly shows that these parallel complete worlds exist in the same space-time. At the same time, the measure density of the particle in different worlds, which can be strictly defined for the discontinuous motion of particle, just provides the objective origin of the measure of different worlds. Thus the visualizing physical picture for many worlds is one kind of subtle time-division existence, in which every world occupies one part of the continuous time flow, and the occupation way is discontinuous in essence, i.e. the time flow for each world is a dense and discontinuous instant set, and all these dense time sub-flows constitute a whole continuous time flow. In this meaning, the many worlds are the most crowded in time!

Although the above many worlds picture of particles or measuring devices can exist in a consistent way, a hard problem does appear when considering the observer, i.e. why does the observer only continuously perceive one definite world while he is still discontinuously moving throughout the many worlds? This seems to be inconsistent with one of our scientific views, according to which our perception is one kind of correct reflection of the objective world. Besides, we must solve the above observer problem in order to have a satisfying many worlds theory. This may need to resort to a theory of consciousness, but we have none up to now.

Now we turn to the second possibility. In the paper[6], a preliminary theory of dynamical collapse is presented based on the discontinuous motion of particles in discrete space-time. Here we will first give some possible evidences for the existence of dynamical collapse, then briefly introduce the theory.

### The discrete space-time and the possible origin of collapse

We have been discussing the motion of particles in continuous space-time, but it should be clearly realized that the continuity of spaced-time is just an assumption. In the nonrelativistic and relativistic domain this assumption can be applicable, and we find no essential inconsistency or paradox. But in the domain of general relativity, the motion of particle and the space-time background are no longer independent, and there exists one kind of subtle dynamical connection between them. Thus the combination of the above evolution law of discontinuous motion (or quantum mechanics) and general relativity may result in essential inconsistency, which requires that the assumption of continuous space-time should be rejected and may further result in the appearance of dynamical collapse. Now let's have a close look at it.

According to general relativity, there exists one kind of dynamical connection between motion and space-time, i.e. on the one hand, space-time is determined by the motion of particles, on the other hand, the motion of particle must be defined in space-time. Then when we consider the superposition state of different positions, say position A and position B, one kind of basic logical inconsistency appears. On the one hand, according to the above evolution law of the discontinuous motion of particles (or quantum mechanics), the valid definition of this superposition requires the existence of a definite space-time structure, in which the position A and position B can be distinguished. On the other hand, according to general relativity, the space-time structure, including the distinguishability of the position A and position B, can't be pre-determined, and it must be dynamically determined by the superposition state of particle. Since the different position states in the superposition state will generate different space-time structures, the space-time structure determined by the superposition state is indefinite. Thus an essential logical inconsistency does appear!

Then what are the direct inferences of the logical inconsistency? First, its appearance indicates that the superposition of different positions of particle can't exist when considering the influence of gravity, since it can't be consistently defined in principle. It should be stressed that this conclusion only relies on the validity of general relativity in the classical domain, and is irrelevant to its validity in the quantum domain. Thus the existence of gravity described by general relativity will result in the invalidity of the linear superposition principle. This may be the physical origin of dynamical collapse of wave function.

Secondly, according to the physical definition of the superposition state of different positions of particle, its existence closely relates to the continuity of space-time, concretely speaking, it requires that the particle in this state should move throughout these different positions during infinitesimal time interval. Thus the nonexistence of this superposition means that infinitesimal time interval based on continuous space-time will be replaced by finite time interval, and accordingly the space-time where the particles move will display some kind of discreteness. In this kind of discrete space-time, the particle can only move throughout the different positions during finite time interval, or we can say, the particle will stay for finite time interval in any position.

Besides, it can prove that when considering both quantum mechanics and general relativity, the minimum measurable time and space size will no longer infinitesimal, but finite Planck time and Planck length. Here we will give a simple operational demonstration. Consider a measurement of the length between points A and B. At point A place a clock with mass $m$ and size $a$ to register time, at point B place a reflection mirror. When $t = 0$ a photon signal is sent from A to B, at point B it is reflected by the mirror and returns to point A. The clock registers the return time. For the classical situation the measured length will be $L = \frac{1}{2}ct$, but when considering quantum mechanics and general relativity, the existence of the clock introduces two kinds of uncertainties to the measured length. The uncertainty resulting from quantum mechanics is: $dL_{QM} \geq (\frac{\hbar L}{mc})^{1/2}$, the uncertainty resulting from general relativity is: $dL_{GR} \geq \frac{Gm}{c^2}$, then the total uncertainty is: $dL = dL_{QM} + dL_{GR} \geq (L \cdot L_p^2)^{1/3}$, where $L_P = (\frac{G\hbar}{c^3})^{1/2}$, is Planck length. Thus we conclude that the minimum measurable length is Planck length $L_P$. In a similar way, we can also work out the minimum measurable time, it is just Planck time $T_P = (\frac{G\hbar}{c^5})^{1/2}$.

Lastly, we want to denote that the existence of discreteness of space-time may also imply that the many worlds theory is not right, and the collapse of wave function does exist. Since there exists a minimal time interval in discrete space-time, and each parallel world must solely occupy one minimal time interval at least, there must exist a maximal number of the parallel worlds during any finite time interval. Then when the number of possible worlds exceeds the maximal number, they will be merged in some way, i.e. the whole wave function will collapse to a smaller state space.

## A theory of dynamical collapse in discrete space-time

In this section, we will briefly introduce the dynamical collapse theory based on the discontinuous motion of particles in discrete space-time. In the paper[6], the discontinuous motion of particles in discrete space-time and its evolution are further analyzed, and a preliminary theory of dynamical collapse in such discrete space-time is presented.

First, the motion state of a particle in discrete space-time is defined. According to the meaning of discrete space-time, the existence of a particle is no longer in one position at one instant as in the continuous space-time, but limited in a space interval $L_P$ during a finite time interval $T_P$. This defines the instantaneous state of particle in discrete space-time. Similar to the situation in continuous space-time, the motion state of particle in discrete space-time is that during a finite time interval much larger than $T_P$, the particle moves throughout the whole space, which proper description is still the measure density $r(x,t)$ and measure flux density $j(x,t)$, but time-averaged. The visual physical picture of such motion will be that during a finite time interval

$T_P$ the particle stays in a local region with size $L_P$, then it will still stay there or appear in another local region, which may be very far from the original region, and during a time interval much larger than $T_P$ the particle will move throughout the whole space with a certain average position measure density $\rho(x,t)$.

After defining the motion state of particle in discrete space-time, the evolution of the discontinuous motion of particles in discrete space-time is analyzed, and a preliminary nonrelativistic evolution equation is worked out. The evolution equation is a revised Schroedinger equation containing two evolution terms, in which the first term is the usual linear Hamiltonian in Schroedinger equation, the second term is a new stochastic nonlinear evolution term resulting from the stochastic change of the position measure density $\rho(x,t)$. It is stressed that the equation is essentially a discrete evolution equation in physics, and all of the quantities are defined relative to the Planck cells $T_P$ and $L_P$. It is further demonstrated that the revised Schroedinger equation will naturally result in the dynamical collapse of wave function. As an example, the dynamical collapse time of a two-level system can be concretely calculated. It is $\tau_c \approx 2k^2 \frac{\hbar E_p}{(\Delta E)^2}$ [1], where $\Delta E$ is the difference of the energy between these two states. Thus the discontinuous motion of particles in discrete space-time turns out to be a possible basis of the theory of dynamical collapse, and may provide a preliminary framework for the complete quantum theory.

In the following, we will further present some possible evidence for the conclusion that the evolution of the discontinuous motion of particles in discrete space-time will naturally result in the dynamical collapse of wave function. First, as to the discontinuous motion of particles in discrete space-time, since the particle does stay in a local region for a finite nonzero time interval, and appears stochastically in another local region during the next time interval, the position measure density $\rho(x,t)$ of the particle, when changed due to the invalidity of the linear superposition principle, will be essentially changed in a stochastic way, which closely relates to the concrete stay time in different stochastic region [1], and the corresponding wave function will be also stochastically changed. Thus the evolution of discontinuous motion in discrete space-time should be the combination of the deterministic linear evolution and stochastic nonlinear evolution.

---

[1] The similar result has also been obtained by Percival[9], Hughston[10] and Fivel[11], and discussed by Adler et al[12].

Secondly, we need to further find the concrete cause resulting in the stochastic change of the position measure density $\rho(x,t)$, and to see whether it will really result in the dynamical collapse of wave function. As we know, the evolution of wave function is determined by the Hamiltonian of the system, or the energy distribution of the system. Thus the stochastic change of the evolution may also relate to the energy distribution of the system. Now consider a simple two-level system, which state is a superposition of two static states with different energy levels $E_1$ and $E_2$, and its position measure density $\rho(x,t)$ will oscillate with the period of $T = \hbar/\Delta E$, where $\Delta E = E_2 - E_1$ is the energy difference. Then if the energy difference $\Delta E$ is so large that it exceeds the Planck energy $E_p$, the position measure density $\rho(x,t)$ will oscillate with a period shorter than the Planck time $T_P$. But as we know, the Planck time $T_P$ is the minimum distinguishable time size in the discrete space-time, and there should be no changes during this minimal time interval. Thus the energy superposition state, in which the energy difference is larger than the Planck energy $E_p$, can't hold all through, and must gradually collapse to one of the energy eigenstates. It can be further inferred that the dynamical collapse process must happen for any energy superposition state due to the general validity of the natural law including the collapse law.

---

[1] As to the discontinuous motion in continuous space-time, the stay time of the particle in any position is zero, thus its position measure density $\rho(x,t)$ is not influenced by the stochastic motion.

The above analysis has indicated that, when the energy difference between different branches of the wave function is large enough, say, for the macroscopic situation[1], the linear spreading of the wave function will be greatly suppressed, and the evolution of the wave function will be dominated by the localizing process. Thus a macroscopic object will be always in a local position, and it can only be still or continuously move in space in appearance. This is just the display of continuous motion in the macroscopic world. Furthermore, it is shown that the evolution law of continuous motion can also be derived from the evolution of the discontinuous motion in discrete space-time[5][6].

## The basis of discontinuous motion

In this section, we will demonstrate that the instant motion of particle should be essentially discontinuous and random. This will give the logical basis of discontinuous motion. Since what quantum mechanics describes should be the discontinuous motion of particles, this may also answer the question 'why the quantum?'.

As we know, the object can move or change its position when there is not any outer cause such as outer force. This is an experiential fact, for example, when you kick a ball, it can move freely afterwards. This fact is well summarized in Newton's first law. Besides, there are also other similar phenomena in the microscopic world, for example, the emission of alpha particles by radioactive isotopes happens without outer cause, or we can say, the alpha particles can spontaneously move out from the radioactive isotopes.

On the other hand, there may exist some deep reasons for this counterintuitive fact. One possible reason is that if the object can't spontaneously move, then the whole world will hold still. As we have known from modern physics, the interaction or force between particles is transferred by the other particles. Now if the particle can't move in a spontaneous way, or it can only move when there is an outer force, then on the one hand, the particle can't move without outer force, on the other hand, the outer force can't exist without the moving particles, which transfer the force.

---

[1] The largeness of the energy difference for a macroscopic object results mainly from the environmental influences such as thermal energy fluctuations.

Thus all particles will be motionless, and all forces will not exist either. In one word, the whole world will be in a deadly still state. The direct inference of this conclusion may be that the world will not exist either. Since there is no motion and interaction, the properties of particle, which closely relates to motion and interaction, will disappear, and the particle devoid of any properties will not exist either. Then the world also disappears, and nothing exists[1]. Thus it seems that the objects must move spontaneously in order to exist.

We can define the ability of the spontaneously moving of object as the nature or activity of object. Then such nature may be taken as the inner cause for the spontaneous motion of object. Since the nature of spontaneous motion of object doesn't change all the while, this kind of inner cause is independent of time and concrete motion processes.

The object can move spontaneously, then how does it move spontaneously? This is a very interesting and important problem. In the following, we will find the instant moving way of the object.

Since the activity of object or the only cause resulting in the spontaneous motion of object is irrelevant to time and concrete motion processes, there is no cause to determine how the object moves spontaneously in space and time. This means that there is no cause to determine the concrete instant motion of the object, i.e. the object is neither determined to move in one special way, nor determined to move in the other special way. Thus the object can only move in a completely random way, or we can say, the instant motion of object must be essentially discontinuous everywhere. As we can see, the reason why the object moves in a random way is just because there isn't any cause to determine a special regular moving way. In short, the object must move, but it doesn't know how to move, so it can only move in a random and discontinuous way.

The above conclusion is also justifiable from a mathematical point of views[5][6]. As we know, the motion state of an object in continuous space-time is the infinitesimal interval state, not the instantaneous state. Then the motion state of the object is a point set in space-time[1], but which type of point set is it? According to the mathematical analysis of point set[2], the natural assumption in logic is that it is a general dense point set in space-time, since we have no *a priori* reason to assume a special form, say a continuous point set. Thus during the infinitesimal interval near any instant the object will always move in a random and discontinuous way.

---

[1] An attractive idea is that the situation in which nothing exists may be logically inconsistent, and can't exist. Then we can interpret the above counterintuitive fact in a complete logical way.

One big obstacle to understand the above conclusion is that people usually think that there exist some laws, say Newton's first law, to determine the existence of a special instantly moving way. Here we will further argue that there doesn't exist such laws at all. Firstly, all laws referring to the time interval, including the infinitesimal time interval, can't determine such instantly moving way. The reason is very simple, since these laws refers to the time-interval motion state of objects, and they are all based on the supposed instantly moving way of objects during the time interval, for example, Newton's first law presupposes the existence of continuous moving way. But in the above discussions, what we consider is the instant motion, not the time-interval state and its evolution, and what we need to find is just the instantly moving way within the time interval. Secondly, physical laws only consider the time-interval motion state of objects. This can be easily seen from the mathematical quantities $dt$ and $dx$ which appear everywhere in physics.

---

[1] Here the point in the point set represents the mass center of the object at one instant.

[2] As we know, the point set theory has been deeply studied since the beginning of the 20th century. Nowadays we can grasp it more easily. According to this theory, we know that the general point set is dense point set, whose basic property is the measure of the point set. While the continuous point set is one kind of special dense point set, and its basic property is the length of the point set. As an example, as to the point set in two-dimensional space-time, the general situation is the dense point set, while the continuous curve is one kind of extremely special dense point set. Surely it is a wonder that so many points bind together to form one continuous curve by order-in fact, the probability for its natural formation is zero.

Furthermore, present physics doesn't analyze the way of instant motion. It only supposes the way of instant motion, for example, classical physics presupposes that the instant motion is continuous.

Now as an example, let's see why the instant motion is not continuous. Since what we analyze is the instant motion, the velocity, which is defined on the time interval, doesn't exist yet, and Newton's first law can't help either. Then the free object has no velocity to hold, and it really doesn't know which direction to move along. Thus the object can't move in a continuous way, since continuous motion requires a definite direction, for example, in one-dimensional situation, the object must select a preferred direction, right or left to move continuously.

### The confirmation of discontinuous motion

It seems very strange that the particles move in a discontinuous and random way. Then can we directly confirm the existence of the discontinuous motion of particles in experiments? In this section, we will demonstrate that protective measurement[7][8] may provide a method to confirm the existence of the discontinuous motion of particles.

As we know, the usual measurement suitable for any motion state will definitely result in the collapse process, thus we can't directly obtain the real motion picture of a single particle using such measurement. Then can we find one kind of new measurement method, which is not suitable for any motion state or completely unknown state, but may not result in the collapse process and can directly reveal the real motion picture of a single particle? The answer is yes, it is just the protective measurement proposed by Aharonov et al[7][8]. In the following, we will first briefly introduce its basic principle.

Protective measurement aims at measure the motion state of a single particle through repeatedly measuring it without destroying its state[1]. By use of this kind of measurement, the motion state or wave function of the particle does not change appreciably when the measurement is being made on it. Its clever way is to let the system undergo a suitable interaction so that it is in a non-degenerate eigenstate of the whole Hamiltonian, then the measurement is made adiabatically so that the motion state or wave function of the particle neither changes appreciably nor becomes entangled with the measurement device. This suitable interaction is called the protection.

---

[1] In real experiment a small ensemble of similar particles may be required.

It should be stressed that the single particle state measured by protective measurement is not completely unknown, we must know before measurement whether the measured system is in the nondegenerate energy eigenstate or which energy superposition state it is in. This is needed for determining whether or how to introduce the protective interaction. Surely, the same protective measurement setting can't measure any motion state. But, all these do not mean that protective measurement is just one kind of meaningless repetition. It does reveal the objective motion state of particle through measuring a single particle, at the same time, even if we must know some information about the measured state of a single particle before measurement, we need not to completely know the state. For example, we only need to know the particle is in the minimum energy eigenstate of some bound potential, which concrete form is unknown, while this state can be naturally achieved through the spontaneous transition of particle.

In the following, we will demonstrate how protective measurement can reveal the discontinuous motion of a single particle, which is described by the position measure density $\rho(x,t)$ and position measure flux density $j(x,t)$, or the complex wave function $\psi(x,t)$. For simplicity but without losing generality, we only consider a particle in a discrete nondegenerate energy eigenstate $\psi(x)$ [1], the interaction Hamiltonian for measuring the value of an observable $A_n$ in this state is: H=g(t)P$A_n$, which couples the system to a measuring device, with coordinate and momentum denoted respectively by Q and P, where $A_n$ is the normalized projection operator on small regions $V_n$ having volume $v_n$, namely:

$$A_n = \begin{cases} \dfrac{1}{v_n}, & x \in V_n; \\ 0, & x \notin V_n. \end{cases} \quad \text{------ (4)}$$

---

[1] For this situation the protection is natural, we need no additional protective interaction. This example has been discussed by Aharonov et al[7][8], here we mainly give its interpretation.

the time-dependent coupling g(t) is normalized to $\int_0^T g(t)dt = 1$, we let g(t) = 1/T for most of the time T and assume that g(t) goes to zero gradually before and after the period T to obtain an adiabatic process when T→∞, the initial state of the pointer is taken to be a Gaussian centered around zero, and the canonical conjugate P is bounded and also a motion constant not only of the interaction Hamiltonian, but of the whole Hamiltonian.

Now using this kind of protective measurement, the measurement of $A_n$ yields the result:

$$<A_n> = \frac{1}{v_n}\int_{v_n} |\psi(x)|^2 dv = |\psi_n|^2 \quad \text{------ (5)}$$

The result $<A_n> = |\psi_n|^2$ is just the average of the position measure density $\rho(x) = |\psi(x)|^2$ over the small region $V_n$, so when $v_n \to 0$ and after performing measurements in sufficiently many regions $V_n$ we can find the position measure density ρ(x) of the discontinuous motion of the measured particle.

Then we will measure the position measure current density j(x) of the discontinuous motion, namely we need measure the value of an observable $B_n = \frac{1}{2i}(A_n \nabla + \nabla A_n)$. The measurement result will be

$$<B_n> = \frac{1}{v_n}\int_{v_n} \frac{1}{2i}(\psi^* \nabla \psi - \psi \nabla \psi^*)dv = \frac{1}{v_n}\int_{v_n} j(x)dv \quad \text{------ (6)}$$

and it is just the average value of the position measure flux density j(x) in the region $V_n$. Then when $v_n \to 0$ and after performing measurements in sufficiently many regions $V_n$, we can also find the position measure flux density j(x) of the discontinuous motion of the measured particle.

Thus we have demonstrated that the discontinuous) motion of a single particle, which is described by the position measure density ρ(x) and position measure flux density j(x), or the complex wave function $\psi(x)$, can be directly confirmed through the above protective measurement. It should be stressed that, in real experiment a small ensemble of similar particles may be required for protective measurement. Besides, we can more easily complete it for the charged particles, for which ρ(x,t) and j(x,t) will represent the effective charge density and current density.

As an example, we briefly analyze the famous double-slit experiment. As we know, the usual position measurement will destroy the double-slit interference pattern, and can't measure the real motion state of the single particle passing through the two slits. Now protective measurement will help to find and confirm the objective discontinuous motion picture of the particle passing through the two slits[13]. According to the principles of protective measurement, since we know the state of the particle beforehand in double-slit experiment, we can protectively measure the objective motion state of the particle when it passes through the two slits. At the same time, the state of the particle will not be destroyed after such protective measurement, and the interference pattern will not be destroyed either. Furthermore, as we have demonstrated above, the results of such protective measurement will show that the position measure density ρ(x,t) of the particle distributes throughout both slits. This will definitely confirm that the particle indeed passes through both slits, and its motion is discontinuous.

## Conclusions

In this paper, we discuss a new realistic interpretation of quantum mechanics based on discontinuous motion of particles. We further give the historical and logical basis of discontinuous motion of particles, and demonstrate that if there exists only one kind of physical reality---particles, then the realistic motion of particles described by quantum mechanics should be the discontinuous motion of particles. It is also shown that protective measurement may provide a direct method to confirm the existence of discontinuous motion of particles.